\documentstyle[amssymb,aps,epsf,twocolumn]{revtex}

\begin{document}
\title{Hypermagnetic Field Effects in the Thermal Bath of Chiral Fermions}
\author{J. Canellos$^{1}$, E. J. Ferrer$^{2}$, and V. de la Incera$^{2}$}
\address{$^{1}$Physics Department, SUNY at Buffalo, Buffalo, NY 14260, USA\\
$^{2}$Physics Department, SUNY-Fredonia, Fredonia, NY 14063, USA}

\address{~\\
\parbox{14cm}{\rm \medskip
The dispersion relations for leptons in the symmetric phase of the
electroweak model in the presence of a constant hypermagnetic
field are investigated. The one-loop fermion self-energies are
calculated in the lowest-Landau-level approximation and used to
show that the hypermagnetic field forbids the generation of the
''effective mass'' found as a pole of the fermions' propagators at
high temperature and zero fields. In the considered approximation
leptons behave as massless particles propagating only along the
direction of the external field. The reported results can be of
interest for the cosmological implications of primordial
hypermagnetic fields. \vskip1mm PACS numbers:11.10Wx, 12.15.Ff,
12.39 Fe, 12.15.Lk, 13.10.+q}}

\maketitle

\narrowtext

{\bf Introduction. }Although the origin of large-scale magnetic
fields found in a number of galaxies, in galactic halos, and in
clusters of galaxies \cite{Galaxies}$,$ remains as an open
question, there are compelling arguments in favor of their
primordial nature. Strong primordial magnetic fields (i.e. fields
of the order of $10^{24}$ $G$ at the electroweak (EW) scale) could
be generated, among others, through different mechanisms during
the electroweak phase transition \cite{EW}$,$ or at the GUT scale
\cite{GUT} (for a recent review on cosmic magnetic fields see e.g.
Ref. \cite{Grasso})$.$ If a primordial magnetic field existed
prior to the EW phase transition only its U(1) gauge component,
i.e. the hypermagnetic field, would survive in the plasma for
infinitely long times, thanks to its long-range character. Its
non-Abelian component, however, would decay because at high
temperatures it acquires an infrared magnetic mass $\sim
g^{2}T$ through its non-linear interactions in the thermal bath \cite{Linde}$%
.$ For Abelian and non-Abelian electric fields a Debye screening
will be always generated by thermal effects producing a
short-range decay for both fields \cite{Fradkin}. Hence, the only
large scale field that can permeate the EW plasma is the
hypermagnetic field.

It has been shown in the past that lepton dispersion relations are
significantly modified by the presence of a constant magnetic
field \cite {Neutrinos}$-$\cite{Ferrer}$.$ However, as it is clear
from the above discussion, in order to consider the effects on
matter of primordial fields before the EW phase transition, the
fermion-hypermagnetic-field interactions should be studied. Notice
that there are essential differences between the interactions of
hypermagnetic and magnetic fields with matter. To mention just
one, we recall that the neutrino, being a neutral particle with
respect to the magnetic field, turns into a hypercharged particle
able to interact with the hypermagnetic field already at the bare
level.

In the symmetric phase of the EW model, fermions are in the
chirally invariant phase with separated left-handed and
right-handed representations of the gauge group. In that phase, a
fermion mass term in the Lagrangian density is forbidden by the
explicit chiral symmetry of the theory. However, as it was found
by Weldon \cite{Weldon}$,$ finite temperature corrections induce a
pole in the fermion Green's function that plays the role of an
effective mass. The induced effective ''mass'' modifies the
fermion dispersion relation in the primeval plasma opening the
door for possible cosmological consequences
\cite{Weldon}$,$\cite{Dolgov}$.$

The main goal of the present paper is to investigate the effects
of a constant hypermagnetic field in the finite-temperature
dispersion relations of leptons in the chiral phase of the EW
model. In our calculations we have only taken into account the
first generation of leptons, but our conclusions can be
straightforwardly extrapolated to the remaining generations. We
find that the hypermagnetic field counteracts the thermal effect
responsible for the creation of effective masses for the chiral
leptons \cite{Weldon}$.$ As a consequence, in the strong field
approximation, where the leptons are basically constrained to the
lowest Landau level (LLL), the chiral leptons behave as massless
particles propagating along the direction of the hypermagnetic
field.

{\bf Fermion Self-Energies in the Ritus' Representation. }To
investigate the fermion dispersion relations in the presence of a
hypermagnetic field we need to find first the finite temperature
fermions' self-energies in the presence of the external field.
From general symmetry arguments (see discussion in \cite{Ferrer}),
one can show that a generic form of the fermion self-energy in
covariant notation can be written as

\begin{equation}
\Sigma (p)=\left[ a_{1}p\llap/_{\Vert }+a_{2}p\llap/_{\bot }+bu%
\llap / +c\widehat{H}\llap/ \right] L  \label{1}
\end{equation}
where $u_{\mu }$ is the four-velocity of the center of mass of the
hypermagnetized medium. In the rest frame where the hypermagnetic
field is defined, we have $u_{\mu }=(1,0,0,0)$ and $u_{\mu }H^{\mu
\nu }=0$ (the strength tensor $H^{\rho \lambda }$ is defined, as
it is customary, in terms of the $U(1)_{Y}$ gauge potential,
$H^{\rho \lambda }=\partial ^{\rho }B^{\lambda }-\partial
^{\lambda }B^{\rho }$). The hypermagnetic field can be written in
covariant form as $H_{\mu }=\frac{1}{2}\varepsilon _{\mu \nu \rho
\lambda }u^{\nu }H^{\rho
\lambda }$. In Eq. (\ref{1}) $%
\widehat{H}_{\mu }=H_{\mu }/\left| H_{\mu }\right| $. The helicity
projector operators are $L,R=\frac{1}{2}%
(1\pm \gamma _{5})$. Notice that to form the structure of $\Sigma
$ in the presence of a hypermagnetic field, we have to consider
the tensor $H_{\mu\nu }$ in addition to the usual tensors $p_{\mu
}$, $u_{\mu }$, $g_{\mu \nu }$, and $\varepsilon _{\mu \nu \rho
\lambda }$, which are always present at finite temperature and
zero field. This enhancement of the tensorial space leads to the
structure with coefficient $c$ in (\ref{1}), and allows to obtain
the separation between longitudinal and transverse momentum terms,
that naturally appears in "magnetic" backgrounds, in a covariant
way

\begin{equation}
p\llap/_{\Vert }=p^{\nu }\widehat{H}_{\nu \rho
}^{*}\widehat{H}^{*\mu \rho }\gamma _{\mu } \label{2}
\end{equation}

\begin{equation}
p\llap/_{\bot }=p^{\nu }\widehat{H}_{\nu \rho }\widehat{H}^{\mu
\rho }\gamma _{\mu } \label{3}
\end{equation}
where $\widehat{H}_{\mu\nu }^{*}=\frac{1}{2\left|
H\right|}\varepsilon _{\mu \nu \rho \lambda }H^{\rho \lambda }$ is
the dual of the dimensionless hypermagnetic field tensor
$\widehat{H}_{\mu\nu }=\frac{1}{\left| H\right|}H_{\mu \nu }$.

The coefficients $a_{1}$, $a_{2}$, $b$, and $c$ in (\ref{1}) are
Lorentz scalars that depend on the parameters of the theory and
the approximation used. Our goal now is to obtain those
coefficients in the one-loop approximation for strong
hypermagnetic fields.

We start from the Schwinger-Dyson equations for the left-handed
(LH) and right-handed (RH) leptons' self-energies in the unbroken
$SU(2)_{L}\otimes U_{Y}(1)$ phase, which are given respectively by

\begin{equation}
\Sigma _{l_{L}}\left( x,y\right) =\Delta _{l_{L}}^{\left(
B,l_{L}\right) }\left( x,y\right) +\Delta _{l_{L}}^{\left(
W^{3,\pm },l_{L}\right) }\left( x,y\right) ,  \label{6}
\end{equation}

\begin{equation}
\Sigma _{e_{R}}\left( x,y\right) =\Delta _{e_{R}}^{\left(
B,e_{R}\right) }\left( x,y\right) .  \label{7}
\end{equation}
Eq. (\ref{6}) denotes, in a generic way, the Schwinger-Dyson
equations for the two LH leptons, $\nu _{L}$ and $e_{L}$. The
$\Delta $'s terms represent the different one-loop bubble diagrams
contributing to the self-energy of the corresponding lepton
(denoted by the subindex in each equation), with the two internal
field lines indicated by the superindices. Their explicit
expressions are

\begin{equation}
\Delta _{l_{L}}^{\left( B,l_{L}\right) }\left( x,y\right) \equiv i\frac{%
g^{\prime \,2}}{4}R\gamma ^{\mu }D_{\mu \nu }^{\left( B\right)
}\left( x-y\right) G_{l_{L}}\left( x,y\right) \gamma ^{\nu }L,
\label{8}
\end{equation}

\begin{equation}
\Delta _{l_{L}}^{\left( W^{3},l_{L}\right) }\left( x,y\right) \equiv i\frac{%
g^{2}}{4}R\gamma ^{\mu }D_{\mu \nu }^{\left( W^{3}\right) }\left(
x-y\right) G_{l_{L}}\left( x,y\right) \gamma ^{\nu }L,  \label{9}
\end{equation}

\begin{equation}
\Delta _{l_{L}}^{\left( W^{\pm },l_{L}\right) }\left( x,y\right) \equiv i%
\frac{g^{2}}{2}R\gamma ^{\mu }D_{\mu \nu }^{\left( W^{\pm }\right)
}(x-y)G_{l_{L}}\left( x,y\right) \gamma ^{\nu }L,  \label{10}
\end{equation}

\begin{equation}
\Delta _{e_{R}}^{\left( B,e_{R}\right) }\left( x,y\right) \equiv
ig^{\prime \,2}L\gamma ^{\mu }D_{\mu \nu }^{\left( B\right)
}\left( x-y\right) G_{e_{R}}\left( x,y\right) \gamma ^{\nu }R.
\label{11}
\end{equation}

In Eqs.(\ref{6})-(\ref{7}) the contributions $\Delta
_{l_{L}}^{\left( \phi ,e_{R}\right) }\left( x,y\right) \equiv
iG_{e}^{2}D_{\phi }\left( x,y\right) G_{e_{R}}\left( x,y\right) $
and $\Delta _{e_{R}}^{\left( \phi ,l_{L}\right) }\left( x,y\right)
\equiv iG_{e}^{2}G_{l_{L}}\left( x,y\right) D_{\phi }(x,y) $ with
internal lines of scalars were neglected, since $G_{e}^{2}\ll
g^{2},g^{\prime \,2}$. Moreover, we do not consider the
contribution of tadpole diagrams, since the early universe, unlike
the dense stellar medium, is almost charge symmetric ($\mu =0$),
with a particle-antiparticle asymmetry of $\sim 10^{-10}-10^{-9}$.

The operators (\ref{8})-(\ref{11}) depend on the LH lepton and RH
lepton propagators, $G_{l_{L}}\left( x,y\right) $ and $G_{e_{R}}\left( x,y\right) $%
, respectively, and on the gauge bosons' propagators $D_{\mu \nu
}\left( x-y\right) $. Neither the Abelian field B, nor the W's,
interact with the external hypermagnetic field, so ignoring group
factors all the gauge bosons' propagators have similar form

\begin{equation}
D_{\mu \nu }\left( x-y\right) =\int \frac{d^{4}q}{\left( 2\pi \right) ^{4}}%
\frac{e^{iq\cdot \left( x-y\right) }}{q^{2}-i\epsilon }\left(
g_{\mu \nu }-\left( 1-\xi \right) \frac{q_{\mu }q_{\nu
}}{q^{2}}\right) ,  \label{12}
\end{equation}
where $\xi $ is the corresponding gauge fixing parameter. The
fermion propagators depend on the hypermagnetic field and cannot
be diagonalized in momentum space with the help of a Fourier
transformation as can be done for the gauge-field propagators
(\ref{12}). Nevertheless, it is possible to extend to the case of
chiral fermions in interaction with a constant hypermagnetic field
a method developed by Ritus \cite{Ritus} to diagonalize the QED
electron self-energy in the presence of a constant electromagnetic
background. Ritus' method has proved to be particularly convenient
to investigate systems under strong magnetic fields \cite{Ng}. In
Ritus' approach a Fourier-like transformation to momentum space of
the fermion Green's function in the presence of a constant
magnetic field is carried out by using transformation functions
(the so called $E_{p}(x)$ functions) which correspond to the
eigenfunctions of the charged particles in the electromagnetic
background. The $E_{p}(x)$ functions play the role, in the
presence of the external background, of the usual Fourier
$e^{ipx}$ functions in the free case. Ritus' method was originally
developed for spin-1/2 charged particles \cite{Ritus}$,$ and it
has been recently extended to the spin-1 charged particle case
\cite{Ferrer}$. $

Using Ritus' approach extended to chiral fermions in a
hypermagnetic background, the left-lepton propagator can be
written as
\begin{equation}
G_{l_{L}}\left( x,y\right) =\sum\limits_{l^{\prime \prime }}\int \frac{%
dk_{0}dk_{2}dk_{3}}{\left( 2\pi \right) ^{4}}E_{k}^{l_{L}}\left(
x\right) \frac{1}{\gamma
.\overline{k}_{l_{L}}}\overline{E}_{k}^{l_{L}}(y)  \label{13}
\end{equation}

where

\begin{equation}
E_{k}^{l_{L}}\left( x\right) =\sum_{\sigma }E_{k\sigma
}^{l_{L}}\left( x\right) \Delta \left( \sigma \right) \label{15}
\end{equation}
with

\begin{eqnarray}
E_{k\sigma }^{l_{L}}\left( x\right) &=&N^{l_{L}}\left( n\right)
e^{i\left(
k_{0}x^{0}+k_{2}x^{2}+k_{3}x^{3}\right) }D_{n}(\rho ),\qquad  \nonumber \\
\rho &=&\sqrt{\left| g^{\prime }H\right| }\left(
x_{1}-2k_{2}/g^{\prime }H\right)  \label{16}
\end{eqnarray}

In Eqs. (\ref{15})-(\ref{16}) the spin-projection matrix $\Delta
\left( \sigma \right)$  and normalization factor $N^{l_{L}}\left(
n\right)$ are respectively given by

\begin{equation}
\Delta \left( \sigma \right) =diag(\delta _{\sigma 1},\delta
_{\sigma -1},\delta _{\sigma 1},\delta _{\sigma -1}),\qquad \sigma
=\pm 1  \label{18}
\end{equation}

\begin{equation}
N^{l_{L}}\left( n\right) =\frac{\left( 4\pi \frac{g^{\prime
}}{2}H\right) ^{1/4}}{\sqrt{n!}} \label{19}
\end{equation}

The values $\sigma =\pm 1$ correspond to the two possible fermion
spin projections. In (\ref{13}) we have considered a constant
hypermagnetic field along the $\mathcal{OZ}$-direction (i.e. an
external
hypermagnetic field with vector potential $B_{\mu }^{ext}=(0,0,Hx_{1},0)$, $%
\mu =0,1,2,3$), and have introduced the notation $\overline{E}%
_{k}^{l_{L}}=\gamma ^{0}(E_{k}^{l_{L}})^{\dagger }\gamma ^{0}$. In
Eq. (\ref{16}) $D_{n}(\rho )$ denotes the parabolic cylinder
functions.

The advantage of the representation (\ref{13}) is that in momentum
space the fermion propagators are simply given in terms of the
eigenvalues

\begin{eqnarray}
\left( \overline{k}_{l_{L}}\right) _{\mu } &=&(k_{0},0,-sgn(g^{\prime }H)%
\sqrt{\left| g^{\prime }H\right| l^{\prime \prime }},k_{3}),\qquad
\nonumber
\\
\qquad \qquad l^{\prime \prime } &=&0,1,2,...  \label{20}
\end{eqnarray}

The integer $l^{\prime \prime }$ in Eq. (\ref{20}) labels the
Landau levels. The relation between the integer $n$, the Landau numbers $%
l^{\prime \prime }$ and the spin projection $\sigma $ is

\begin{eqnarray}
n=n(l^{\prime \prime },\sigma )\equiv l^{\prime \prime }+sgn(g^{\prime }H)%
\frac{\sigma }{2}-\frac{1}{2},\qquad n &=&0,1,2,...  \label{22}
\end{eqnarray}

The corresponding formulae for the right-electron Green's function
$G_{e_{R}}\left( x,y\right) $ are straightforwardly obtained from
(\ref{13})-(\ref{19}) with the change of notation $l_L\to e_R$ and
the replacement of $g^{\prime }$ by $2g^{\prime }$.

The transformation to momentum space of the left leptons'
self-energies (\ref{6}) can be done with the help of the functions
(\ref{15}) in the following way:

\begin{equation}
(2\pi )^{4}\delta ^{(4)}(p-{p}^{\prime })\Sigma
_{l_{L}}(\overline{p})=\int
d^{4}xd^{4}y\overline{E}_{p}^{l_{L}}(x)\Sigma
_{l_{L}}(x,y)E_{p^{\prime }}^{l_{L}}(y)  \label{23}
\end{equation}
Similarly, for the right-electron self-energy we have,

\begin{equation}
(2\pi )^{4}\delta ^{(4)}(p-{p}^{\prime })\Sigma
_{e_{R}}(\overline{p})=\int
d^{4}xd^{4}y\overline{E}_{p}^{e_{R}}(x)\Sigma
_{e_{R}}(x,y)E_{p^{\prime }}^{e_{R}}(y)  \label{24}
\end{equation}
In (\ref{23})-(\ref{24}) we introduced the notation $\delta ^{(4)}(p-{p}%
^{\prime })=\delta _{ll^{\prime }}\delta (p_{0}-p_{0}^{\prime
})\delta (p_{2}-p_{2}^{\prime })\delta (p_{3}-p_{3}^{\prime })$.

Let us consider now the Schwinger-Dyson equation for the left-lepton (\ref{6}%
) in the one-loop approximation. With this aim, we substitute the
left-fermion propagators (\ref{13}) and the gauge boson propagators (\ref{12}%
) (taken in the Feynman gauge $\xi =1)$ in Eq. (\ref{6}).
Transforming it to momentum space, and using Eq. (\ref{23}), we
obtain \[ (2\pi )^{4}\delta ^{(4)}(p-{p}^{\prime })\Sigma
_{l_{L}}(\overline{p})=\]

\begin{eqnarray}
&=&iG_{l_{L}}^{2}R\int d^{4}xd^{4}x^{\prime }\sum_{l^{\prime
\prime
}=0}^{\infty }\int \frac{d^{3}k}{\left( 2\pi \right) ^{4}}\int \frac{d^{4}q}{%
\left( 2\pi \right) ^{4}}\frac{e^{iq\cdot \left( x-x^{\prime }\right) }}{%
q^{2}}  \label{24-a} \\
&&\times \bar{E}_{p}(\frac{g^{\prime }}{2},x)\gamma ^{\mu }E_{k}(\frac{%
g^{\prime }}{2},x)\left( \frac{1}{\gamma \cdot \overline{k}}\right) \bar{E}%
_{k}(\frac{g^{\prime }}{2},x^{\prime })\gamma _{\mu }E_{p^{\prime }}(\frac{%
g^{\prime }}{2},x^{\prime })L.\nonumber
\end{eqnarray}

In (\ref{24-a}) we introduced the notation
$G_{l_{L}}^{2}=[\frac{(g^{\prime })^{2}}{4}+\frac{3(g)^{2}}{4}]$.
Integrating in $x$, $y$, and $k,$ and doing
the spin sums, the left-lepton self energy takes the form\footnote{%
Detailed calculations will be published elsewhere.}

\[
\sum\nolimits_{l_{L}}(\overline{p})=-2iG_{l_{L}}^{2}\sum\limits_{l^{^{\prime
\prime }}=0}^{\infty }\int \frac{d^{4}q}{(2\pi )^{4}}\frac{e^{-\frac{1}{%
2\left| g^{\prime }H\right| }q_{\bot
}^{2}}}{q^{2}}\frac{1}{\overline{\kappa }^{2}}
\]

\[
\times \left\{ \delta _{l,l^{^{\prime \prime }}}(\gamma ^{\bot }\overline{%
\kappa }_{\bot })+(\delta _{l,l^{^{\prime \prime }}-sgn(g^{\prime
}H)}\Delta (1)\right.
\]

\begin{equation}
+\left. \delta _{l,l^{^{\prime \prime }}+sgn(g^{\prime }H)}\Delta
(-1)\right) \left. (\gamma ^{\Vert }\overline{\kappa }_{\Vert
})\right\} L. \label{24-b}
\end{equation}
where $\overline{\kappa }_{\mu }=(p_{0}-q_{0},0,-sgn(g^{\prime }H)\sqrt{%
\left| g^{\prime }H\right| l^{^{\prime \prime }}},p_{3}-q_{3})$.

At this point we assume that the hypermagnetic field is strong
enough as to use the LLL approximation. To justify such an
approximation for cosmological applications we should recall that
due to the equipartition principle, the magnetic energy can only
be a small fraction of the universe energy density.
This argument leads to the relation between field and temperature $%
H/T^{2}\sim 2$ \cite{Shaposhn}. For such fields, the effective gap
between the Landau levels (LL) is $g^{\prime }H/T^{2}\sim {\cal
O}(1)$. In this case the weak-field approximation (where the sum
in all LL is important), cannot be used because field and
temperature are comparable. On the other hand, because the thermal
energy is of the same order of the energy gap between LL's, it is
barely enough to induce the occupation of just a few of the lowest
Landau levels. Therefore, it is natural to expect that the LLL
approximation will provide a good qualitative description of the
leptons' propagation in the presence of strong hypermagnetic
fields, although clearly a more quantitative treatment of the
problem in a field $H\sim 2T^{2}$ would require numerical
calculations due to the lack of a leading parameter.

It can be seen from Eq. (\ref{24-b}) that in the LLL
approximation, on which the external Landau number $l=0$, the term
proportional to $\delta _{l,l^{^{\prime \prime }}}$ vanishes because $\overline{\kappa }%
_{\bot }=(0,0,-sgn(g^{\prime }H)\sqrt{\left| g^{\prime }H\right|
l^{^{\prime \prime }}},0)=0$. Depending on the sign of $g^{\prime
}H$, only one of the other two terms can survive. Assuming
$g^{\prime }H >0$, the surviving term is the one proportional to
$\Delta \left( 1\right)$, for which the internal Landau
number is $l^{\prime \prime }=1$. Therefore, just one spin projection ($%
\sigma = 1$ in this case) contributes to the self energy in the
LLL. It is worthwhile to note that the LLL approximation also
provides a field-dependent infrared cutoff in the denominator of
Eq. (\ref{24-b}), since for $l^{^{\prime \prime }}=1$ one has $\overline{\kappa }%
^{2}=-(p_{0}-q_{0})^{2}+\left| g^{\prime }H\right|
+(p_{3}-q_{3})^{2}$.

Taking into account the above considerations, in the LLL approximation Eq. (%
\ref{24-b}) reduces to
\begin{equation}
\Sigma _{l_{L}}(\overline{p})=I_{l_{L}}(\bar{p})(-i\gamma
_{4}-\gamma _{3})L,\quad \quad \quad  \label{25}
\end{equation}
where the Wick rotation to Euclidean space $(q_{0}\rightarrow
iq_{4},\;p_{0}\rightarrow ip_{4},\gamma _{0}\rightarrow i\gamma
_{4})$ has been performed, and the coefficient
$I_{l_{L}}(\bar{p})$ is given by

\[
I_{l_{L}}(\bar{p})=2G_{l_{L}}^{2}\int \frac{d^{4}q}{\left( 2\pi \right) ^{4}}%
\frac{\exp -(\frac{1}{\left| g^{\prime }H\right| }q_{\bot }^{2})}{%
q_{4}^{2}+q_{\bot }^{2}+q_{3}^{2}}
\]

\begin{equation}
\times \frac{i\left( \bar{p}_{4}-q_{4}\right) -\left( \bar{p}%
_{3}-q_{3}\right) }{\left( \bar{p}_{4}-q_{4}\right) ^{2}+\left|
g^{\prime }H\right| +\left( \bar{p}_{3}-q_{3}\right) ^{2}}
\label{27}
\end{equation}

In a similar way, one can show that in the LLL-approximation the
one-loop right-lepton self energy takes the form

\begin{equation}
\Sigma _{e_{R}}(\overline{p})=I_{e_{R}}(\bar{p})(-i\gamma
_{4}+\gamma _{3})R, \label{26}
\end{equation}
with $I_{e_{R}}(\bar{p})$ given as

\[
I_{e_{R}}(\bar{p})=2(g^{\prime })^{2}\int \frac{d^{4}q}{\left(
2\pi \right)
^{4}}\frac{\exp -(\frac{1}{2\left| g^{\prime }H\right| }q_{\bot }^{2})}{%
q_{4}^{2}+q_{\bot }^{2}+q_{3}^{2}}
\]

\begin{equation}
\times \frac{i\left( \bar{p}_{4}-q_{4}\right) +\left( \bar{p}%
_{3}-q_{3}\right) }{\left( \bar{p}_{4}-q_{4}\right) ^{2}+2\left|
g^{\prime }H\right| +\left( \bar{p}_{3}-q_{3}\right) ^{2}}
\label{28}
\end{equation}

As mentioned above, the $H$-dependent term in the denominators of
(\ref{27}) and (\ref{28}) acts as an effective infrared cutoff. In
this sense, a strong hypermagnetic field can cure the infrared
divergences of the massless chiral fermions in the symmetric phase
of the EW model.

The coefficients $I_{l_{L}}(\bar{p})$ and $I_{e_{R}}(\bar{p})$ are
now calculated at finite temperature with the help of the
Matsubara formalism. This can be done as usual by substituting
$\int d^{4}q\rightarrow \sum\limits_{q_{4}}\int d^{3}q$, in
(\ref{27}) and (\ref{28}), and then summing over discrete
frequencies, corresponding in our case to the boson four-momentum
$q_{4}=2n\pi /\beta $, $(n=0,1,2,...)$, with $\beta $ being the
inverse of the temperature $\beta =1/T$. After summing in $n$ we
obtain

\begin{equation}
I_{l_{L}}(\bar{p})=I_{l_{L}}^{(vac)}(\bar{p})+I_{l_{L}}^{\beta
}(\bar{p}) \label{29}
\end{equation}

\begin{equation}
I_{e_{R}}(\bar{p})=I_{e_{R}}^{(vac)}(\bar{p})+I_{e_{R}}^{\beta
}(\bar{p}) \label{30}
\end{equation}
where the vacuum ($I_{l_{L}}^{(vac)}$, $I_{e_{R}}^{(vac)})$ and the thermal ($%
I_{l_{L}}^{\beta }$, $I_{e_{R}}^{\beta }$) parts can be
transformed back to Minkowski space to be expressed for small
momenta ($\sqrt{H}\gg \left| p_{_{\Vert }}\right| $) as

\begin{equation}
I_{l_{L}}^{(vac)}(\bar{p})=\frac{G_{l_{L}}^{2}}{8\pi ^{2}}a\left(
p_{0}-p_{3}\right)  \label{31}
\end{equation}

\begin{equation}
I_{e_{R}}^{(vac)}(\bar{p})=\frac{(g^{\prime })^{2}}{8\pi
^{2}}a\left( p_{0}+p_{3}\right)  \label{32}
\end{equation}

\begin{equation}
I_{l_{L}}^{\beta }(\bar{p})=\frac{G_{l_{L}}^{2}}{2\pi }\left[ \frac{T^{2}}{%
\left| g^{\prime }H\right| }\right] \left( p_{0}-p_{3}\right)
\label{33}
\end{equation}

\begin{equation}
I_{e_{R}}^{\beta }(\bar{p})=\frac{(g^{\prime })^{2}}{2\pi }\left[ \frac{T^{2}%
}{2\left| g^{\prime }H\right| }\right] \left( p_{0}+p_{3}\right)
\label{34}
\end{equation}
with $a$ being a positive constant of order one ($a\simeq 0.855$).

{\bf Dispersion Relations.} Using the results
(\ref{29})-(\ref{34}), the structures of the LH and RH
self-energies can be written as

\begin{equation}
\Sigma _{l_{L}}(\overline{p})=\left[ A_{L}p\llap/_{\shortparallel
}-A_{L}\,p^{\mu }\widehat{H^{*}}_{\mu \nu }\gamma ^{\nu }\right] L
\label{35}
\end{equation}

\begin{equation}
\Sigma _{e_{R}}(\overline{p})=\left[ A_{R}p\llap/_{\shortparallel
}+A_{R}p^{\mu }\widehat{H^{*}}_{\mu \nu }\gamma ^{\nu }\right] R
\label{36}
\end{equation}
with

\begin{equation}
A_{L}=\frac{G_{l_{L}}^{2}}{2\pi }\left[ \frac{a}{4\pi
}+\frac{T^{2}}{\left|
g^{\prime }H\right| }\right] ,\qquad A_{R}=\frac{(g^{\prime })^{2}}{2\pi }%
\left[ \frac{a}{4\pi }+\frac{T^{2}}{2\left| g^{\prime }H\right|
}\right] \label{37}
\end{equation}

Notice that at strong field the structure appearing in
Eq.(\ref{1}) with coefficient $a_{2}$ is absent because
$p\llap/_{\bot }=0$ in the LLL approximation. Such a term was not
present either in the case of neutrinos in strong magnetic field
Ref.\cite{Ferrer}, since there $a_{2}=0$.

It is easy to realize that the second structures in Eqs.
(\ref{35}) and (\ref {36}) are nothing but the sum of the last two
structures of (\ref{1}), with coefficients $b$ and $c$ for left
and right leptons given respectively by
\begin{eqnarray}
b_{L} &=&-A_{L}(p\cdot H)  \label{38-a} \\
b_{R} &=&A_{R}(p\cdot H)  \label{38-b}
\end{eqnarray}
and
\begin{eqnarray}
c_{L} &=&A_{L}(p\cdot u)  \label{39-a} \\
c_{R} &=&-A_{R}(p\cdot u)  \label{39-b}
\end{eqnarray}

The fact that the two structures in (\ref{35}), as well as in
(\ref{36}), have the same coefficient will have significant
consequences for the propagation of chiral leptons in the
hypermagnetized medium, as one can immediately corroborate by
solving the lepton dispersion equations.

In general, the fermion dispersion equation including radiative
corrections is given by

\begin{equation}
\det \left[ \overline{p}\cdot \gamma +\Sigma \right] =0 \label{39}
\end{equation}
Substituting $\Sigma $ with the expressions (\ref{35})-(\ref{36}),
and taking the determinant, we obtain a similar equation for the
LH and RH leptons in the LLL approximation

\begin{equation}
-(1+2A_{L,R})p_{0}^{2}+(1+2A_{L,R})p_{3}^{2}=0  \label{40}
\end{equation}
This result indicates that in a strong hypermagnetic field the
chiral fermions behave as massless particles which propagate only
along the longitudinal direction ($p_{0}^{2}=p_{3}^{2}$).

{\bf Conclusions.} In the present paper we calculate the chiral
leptons' dispersion relations in the presence of a strong
hypermagnetic field. We emphasize that although the one-loop
self-energies for LH and RH leptons have different contributions
(see Eqs. (\ref{6})-(\ref {7})), their dispersion relations in the
LLL approximation coincide. A main outcome of our investigation is
that the hypermagnetic field counteracts the temperature effects
on the chiral leptons' dispersion relations. As mentioned above,
according to Weldon\cite{Weldon} the dispersion relations of the
chiral leptons are modified due to high-temperature effects in
such a way that a temperature-dependent pole appears in their
one-loop Green's functions (i.e. an effective mass). As we have
shown above, at sufficiently strong hypermagnetic fields such a
pole disappears. This effect can be of interest for cosmology. In
particular, if a primordial hypermagnetic field could exist in the
EW plasma prior to the symmetry-breaking phase transition, with
such a strength that the leptons would be mainly confined to the
LLL for the existing temperatures, then the thermal effective mass
found for the leptons in that phase at zero field\cite {Weldon}
will be swept away. We expect that for moderate field strengths
$H\ll $$T^{2}$, the leptons' effective mass will be different from
zero, but smaller than the one found at zero field\cite{Weldon}$.$

Another important outcome of our investigation is the large
anisotropy in the propagation of leptons, since in the strong
hypermagnetic field domain leptons are mainly restricted to move
along the direction of the field lines. In particular, if one
compares the effects of high magnetic fields on neutrino
propagation\cite{Ferrer}$,$with those produced by a strong
hypermagnetic field, one can notice that while both fields give
rise to anisotropic propagations, the magnetic field allows
neutrino modes of propagation in all directions, but the
hypermagnetic field restricts the motion only to the longitudinal
direction. This different behavior can be understood from the fact
that the neutrino is minimally coupled to the hypermagnetic field
because of its nonzero hypercharge, but since it is electrically
neutral, it can couple to the magnetic field only through
radiative corrections.

From the above discussion, one can envision that if large
hypermagnetic fields were present in the unbroken EW\ phase of the
early universe, the propagation of leptons in general and
neutrinos in particular would be highly anisotropic, an effect
that could leave a footprint in a yet to be detected relic
neutrino cosmic background. If such a footprint were observed, it
would provide a direct experimental proof for the existence of
strong hypermagnetic fields in the early universe.

\begin{acknowledgements}
The work of E.J.F. and V.I. was supported in part by the National
Science Foundation under Grant No. PHY-0070986.
\end{acknowledgements}

\end{document}